\documentclass{article}

\usepackage{arxiv2}

\usepackage[utf8]{inputenc} 
\usepackage[T1]{fontenc}    
\usepackage{hyperref}       
\usepackage{url}            
\usepackage{booktabs}       
\usepackage{amsfonts}       
\usepackage{nicefrac}       
\usepackage{microtype}      
\usepackage{lipsum}
\usepackage{graphicx}
\graphicspath{ {./images/} }

\usepackage{pagenote}
\usepackage{enumerate}

\usepackage{url}

\title{A new degree of freedom for opinion dynamics models: the arbitrariness of scales}

\author{
Dino Carpentras\textsuperscript{1,2}
  \And
Alejandro Dinkelberg\textsuperscript{1,2}
 \And
Michael Quayle\textsuperscript{1,3}
  \And
  \\
\textsuperscript{1}Social Dynamics Lab, Department of Psychology, \\ Centre for Social Issues Research, University of Limerick, Ireland \\
\textsuperscript{2}MACSI (Mathematics Applications Consortium for Science and Industry), \\ Department of Mathematics and Statistics,  University of Limerick, Ireland \\
\textsuperscript{3}Department of Psychology, School of Applied Human Sciences,\\ University of KwaZulu‐Natal, Scottsville, South Africa
}

\usepackage{natbib}
	\setcitestyle{authoryear,round,aysep={}}
	
\begin{document}
\maketitle
\begin{abstract}
Opinion dynamics models have been developed to study and predict the evolution of public opinion. Intensive research has been carried out on these models, especially exploring the different rules and topologies, which can be considered two degrees of freedom of these models. In this paper we introduce what can be considered a third degree of freedom.

Since it is not possible to directly access someone's opinions without measuring them, we always need to choose a way to transform real world opinions (e.g. being anti-Trump) into numbers. However, the properties of this transformation are usually not discussed in opinion dynamics literature. For example, it would be fundamental to know if this transformation of opinions into numbers should be unique, or if several are possible; and in the latter case, how the choice of the scale would affect the model dynamics.

In this article we explore this question by using the knowledge developed in psychometrics. This field has been studying how to transform psychological constructs (such as opinions) into numbers for more than 100 years. 

We start by introducing this phenomenon by looking at a simple example in opinion dynamics. Then we provide the necessary mathematical background and analyze three opinion dynamics models introduced by Hegselmann and Krause. Finally, we test the models using agent-based simulations both in the case of perfect scales (infinite precision) and in the case of real world scales. 

Both in the theoretical analysis and in the simulations, we show how the choice of the scale (even in the case of perfect accuracy and precision) can strongly change the model’s dynamics. Indeed, by choosing a different scale it is possible to (1) find different numbers of final opinion clusters, (2) change the mean value of the final opinion distribution up to a change of $\pm 100 \%$ and (3) even transform one model into another.

We conclude our analysis by discussing how the scale acts as a new (and still unexplored) degree of freedom for opinion dynamics models. This result affects both researchers interested in exploring the properties of different models and those interested in using them in practical cases. Exploration of its effects on different models is a key step in understanding the properties of these models and how they could be successfully employed in real-world applications. Thus, in our conclusion, we list a series of possible studies which will expand our knowledge on this degree of freedom and speed up the process of connecting the models to real-world data.
\end{abstract}



\keywords{opinion dynamics \and agent based models \and agent based simulations \and empirical validation \and degrees of freedom \and scale \and psychometrics}





\section{Introduction}

\subsection{Opinion dynamics and its degrees of freedom}
In recent years, more and more opinion dynamics models have been developed and studied. These models use a population of interacting entities, usually called \textit{agents}. Each agent has at least one variable representing her own opinion \citep{Castellano_2009, Flache}. The opinion has been represented as a continuous, discrete (ordinal) or nominal variable, both in monodimensional and multidimensional spaces. The space in which an opinion can be found (usually a subset of the real numbers) is referred to as the \textit{opinion space}. For the rest of this paper, we will focus on one of the most popular choices, which is the case of continuous monodimensional opinions. For example, representing an agent's opinion in a continuous opinion space ranging between zero and one. 

Some attempts have been made to generalize these models, especially to understand what makes different formalisations similar or events identical \citep{Coates}. At the moment, the two main degrees of freedom are the \textit{interaction rule} and the \textit{network topology}.

In the \textit{network}, nodes represent the agents while edges represent some kind of connection between agents. Usually if no edge exists between two agents it is not possible for those agents to interact, whereas if an edge exists, interaction between these agents is possible (depending on the interaction rule). The \textit{interaction rule}, specifies how agents interact with each other. That is, how they choose partners for the interaction and how they update their opinion after the interaction has taken place. In this article, we show that the decisions made about how to represent psychological constructs as numbers constitute an additional degree of freedom.

\subsection{Representing opinions using numbers}
Representing opinions as ordered numbers has become common in our society \citep{Danziger1994}. Psychometric scales are routinely used to convert a construct (such as someone’s opinion or intelligence) into numbers. Indeed, it is common to assume a correspondence between psychometric measures and numbers in opinion dynamics models. The assumption is general enough that usually authors do not need to discuss the properties of the model’s opinion space \citep{Castellano_2009, Flache}, or can directly use psychological data as input for their models \citep{Duggins, Jia, Innes}. 

A fundamental feature of psychometric scales is that the choice of the scale itself is an arbitrary step, meaning that is possible to find several different scales which are equally valid for measuring the same construct. A similar situation can be seen in physics, where usually we do not have any criterion for claiming that one unit is better than another (e.g. meters vs feet). However, in physics this process is generally limited to linear transformation of the same units (e.g. $ 1 m = 3.3 ft$), while psychological scales also admit nonlinear transformations. For example, by squaring a unit of distance, we will obtain a unit of area. However, in psychometrics if $x$ is a scale measuring opinions on vaccination,  $x^2$ is considered a scale measuring the same construct on a transformed axis. Indeed, in empirical social science research it is recommended practice to re-scale psychometric data using nonlinear functions (e.g. using $x^2$ or exponential transformation when compensating for skewness)\citep{Tabachnick, Howell2012}.

Another way to investigate this peculiarity of scales, comes from the fact that in in the physical world we can combine two objects to obtain a new one. Thus, we can have two rods of the same length and, after combining them, we obtain a rod twice as long as the original one. Indeed, we can say that a 2 meters object is twice as long as a 1 meter one. This operation is not possible with opinions; it is not possible to combine two identical opinions to obtain an opinion twice as strong as the original ones. Consequently, if person A scored 2 and person B scored 4 on an opinion scale, we cannot say that B’s opinion is twice as strong as A’s one. However, these scales are still giving us some kind of information. We can still say that B's opinion is stronger than A's one. All the scales that preserve this order (i.e. A's opinion < B's opinion) are commonly referred to as \textit{ordinal scales}\footnote{Here we assumed that ordinality is preserved \textit{across} participants. This is often referred to as inter-individual measurement invariance. If this does not exist, then it is possible that measures are still ordinal but \textit{within} participants (e.g. the same participant at two time points).}.

Notice that is also possible to have scales which are not ordinal. For example, if we are supposed to distinguish between people who prefer basketball, baseball or football we would use what is called an \textit{nominal scale}. Indeed, it does not make sense to establish a hierarchy between these categories (e.g. people who prefer basketball > people who prefer baseball).

Since the only requirement for ordinal scales is to preserve order, when these scales undergo transformations they will still be ordinal scales (as long as the transformation is an order preserving one). For example, $2<3$ and $2^2<3^2$. Thus it is possible to use both linear transformations and non linear transformations such as $x^2$, square root, logarithmic, exponential, etc \footnote{Note that some transformations, such as $x^2$, preserve order only when the scores are positive.}. Indeed, most of them are routinely used for data transformation before applying linear models \citep{Tabachnick, Howell2012}.

Another brief example of how we can choose different ordinal scales for the same opinion comes from the use of Likert-type scales. These scales are obtained from questionnaires which present a statement and offer a group of possible response options, out of which one is chosen by the participant (e.g. Bad, Neutral, Good). Usually, response options are coded as evenly spaced natural numbers for analysis (i.e. [1, 2, 3] for [Bad, Neutral, Good]). The anchor points are at the discretion of the researcher and are frequently symmetric around a presumed midpoint. For example, Wada et al. \citep{Wada} use the response options [Very good, Good, Somewhat good, Somewhat bad, Bad, and Very bad] coded numerically as [1, 2, 3, 4, 5, 6]. 

As this is an ordinal scale, it is neither clear that the response options "good" and "bad" are equidistant from the presumed midpoint nor that the modifiers "very" and "somewhat" have similar weight in conjunction with "good" and "bad." This question was explored in a study by YouGov \citep{YouGov} where people were asked to rate a large number of common response options (e.g. "Good" or "Bad"). Specifically, people rated how much positive each option was in an interval ranging from 0 (extremely negative) to 10 (extremely positive). 

\begin{figure}[!t]
\centering
\includegraphics[width=0.8\textwidth]{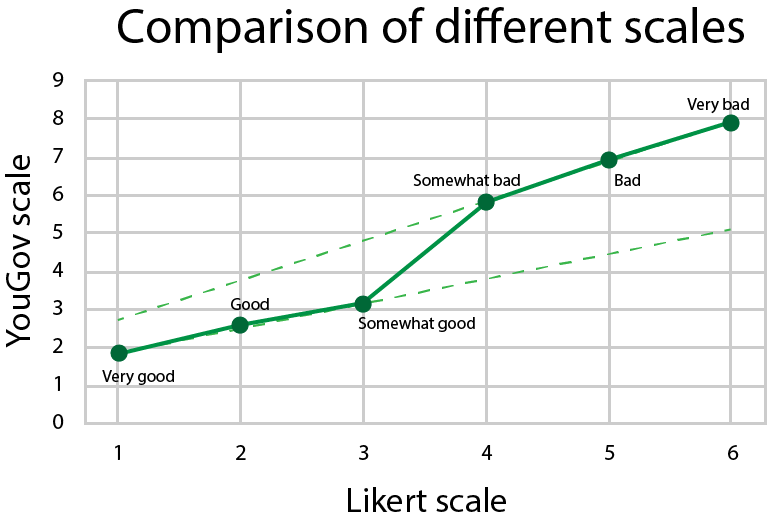}
\caption{Likert scale obtained turning the answers [Very good, Good, Somewhat good, Somewhat bad, Bad, very bad] into the numbers [1, 2, 3, 4, 5, 6] as used in Wada et al. These values have been plotted against the respective values from the YouGov study. The graph shows the lack of linearity between the two scales, as well as the preservation of the order.}
\label{fig:X10}
\end{figure}

It is then possible to use these values to see how they compare with the values used in Wada’s Likert scale. The result is shown in figure \ref{fig:X10}, where we can clearly see both the ordinality of the two scales (the bigger the $x$ values, the bigger the $y$ values) and the lack of linearity between the positive and negative portions of the response scale.

Note also that the YouGov study showed asymmetry in the expressions which are commonly used as semantically  symmetric response options in Likert scales, such as Good (avg. 6.92) vs Bad (2.60) or very bad (1.76) vs very good (7.90). This asymmetry can also be seen in the graph as the slope obtained from the first three answers is different to the one from the last three (see the dashed lines).

A special case of ordinal scales is the category of \textit{interval scales} which have the same properties of the units of measurements used in physics. Indeed, units of measurement from physics are interval scales (within the framework of measurement theory) \footnote{Here we do not discuss ratio scales, which are a special case of interval scales. This is done because ratio scales are not theoretically possible within psychometrics.}. Using interval scale it is possible to say if opinion A is stronger than opinion B (as in ordinal scales), and also how \textit{much} stronger it is (e.g. A's opinion is twice as strong as B's one). Indeed, different interval scales can be obtained from each other via linear transformation (in the case of zero noise) in the same way as how we convert meters to feet. 

Interval scales were first put forward by Stevens \citep{Stevens46, Stevens51}, and Luce and Tukey show that they are theoretically obtainable for psychometric measurements \citep{Luce64}. However it is still not clear whether it is possible to ever obtain this level of measurement precision, especially in real-world research settings \citep{Borsboom, Kyngdon, Keats, Domingue}. Certainly, the requirements for ensuring interval-level psychometric measurement are very stringent indeed and rarely met in psychological data \citep{Heene, Borsboom}. Therefore ordinal scales are the most common level of measurement for opinion representation. 

To sum up this section, the choice of a scale that represents a specific opinion is arbitrary. That is, if $S$ is a scale that can be used for measuring opinions, then $S^2$ and $e^S$ are also equally valid ordinal scales. In the extremely rare case involving interval scales, we could either limit our analyses only to linear transformations (to stay in the same category), or we could apply nonlinear transformations and obtain ordinal scales. In the next sections we will show how this possibility of choosing different scales for the same opinion can be represented as a new degree of freedom in opinion dynamics models.

\subsection{An example of opinion prediction}

To better understand how the choice of scale may impact predictions of opinion dynamics models, let us consider first a naive example. Here we consider two people with two different opinions on the same subject, say one with a pro-Trump opinion and the other with an anti-Trump opinion. Several models (such as the Deffuant model \citep{Deffuant2000} or the Hegselmann Krause model \citep{HK2002}) assume that if two people interact several times, their opinions will converge to an average value (let us neglect the confidence interval for this example). This means that they may end up being either both pro-Trump, both neutral or both anti-Trump, depending on their starting points.

Thus, we can consider an opinion space in the interval [0,1]. We choose 0.5 to represent the neutral point separating the anti-Trump opinion (values < 0.5) from the pro-Trump opinion (values > 0.5). Supposing two individuals have opinions of 0.3 and 0.9, the model predicts that they will converge to the opinion represented as 0.6 (obtained as the average between the two), thus, a pro-Trump opinion.

As previously mentioned, the representation of opinions as numbers is arbitrary and in general is not symmetric (showed in the example above regarding the distance of \textit{good} and \textit{bad} from the center). So what happens if we use different numerical values to represent the same opinions? When measuring the same people on a different scale, the same two opinions may be represented as 0.4 and 0.6 (still within the interval [0,1]). 

Note here that obtaining different numbers representing the same constructs could occur for many reasons. For example if, in one scale, a transformation was applied to compensate for skewness. Or perhaps different phrasing is used for response options; or questions are phrased differently; or one scale uses an odd number of response options and the other uses an even number \citep{taherdoost}. 

In any case, if in the second scale represented the opinions as 0.4 and 0.6, the model will predict convergence to neutrality (average = 0.5). Considering yet another scale, these same opinions can be represented as 0.2 and 0.6 which predicts a convergence to an anti-Trump opinion, the opposite to what was predicted when using the first scale.

Of course, this example is arbitrary and superficial, and it should not be considered as proof. Nonetheless, it allows us to see how the choice of the scale may affect the dynamics of a model. Furthermore, it pushes us to ask several related questions: is this phenomenon due to noise? Or is it negligible when using many more agents? Does anything change if we include the confidence interval? In the next sections we will address these questions showing how the choice of the scale naturally appears as a new degree of freedom in opinion dynamics models.

Specifically, in section 2 we clarify the concept of scale, to eliminate ambiguity in the following sections. In section 3 we introduce the mathematical concept of \textit{topological conjugacy}, which will be used to test if two models are the same. In section 4 we will analyze mathematically different models, especially focusing our attention on three models introduced by Hegselmann and Krause. In this section we will also show that these three models are the same under scale transformation. In section 5 we will repeat our analysis of section 4 using agent based simulations, thus validating previous results and shining more light on their meaning. In section 6 we will test the same models using real world scales and showing how the choice of scale may similarly change the model predictions. Finally, in section 7 we will summarize our findings and suggest possible follow up studies to further explore the importance of the choice as scale as a degree of freedom in opinion dynamics models.


\section{Definitions and distinctions}
\subsection{Construct}

In the previous sections we used the term \textit{construct} to represent abstract entities such as opinions or intelligence which can be measured using scales. Constructs are usually considered to be abstract concepts, thus not something physical within someone's brain \citep{cronbach1955construct}. As we are focused on the field of opinion dynamics, in this text we will use the terms \textit{opinions} and \textit{constructs} interchangeably. However, even in this field, some models use other parameters besides opinion, such as stubbornness \citep{Friedkin}, quality of information \citep{Brooks} and many others \citep{Duggins}. Thus, our findings are not limited to opinions, but to every construct represented as a continuous variable in opinion dynamics models.

It is also important to distinguish between a construct and its representation into a scale. The two terms are often confused as we tend to think of a construct by its numerical value. In the physical world, this would be the difference between the length of an object (which is a property of the object itself) and the representation of that length into a scale (e.g. 1 m or 3.3 ft). The fundamental distinction is that changing a scale does not alter the properties of the object itself (i.e. the object does not become longer or shorter when we change units). In the same way, when discussing the following sections, we should keep in mind the difference between someone’s opinion and its representation on a specific scale.

Since it is not possible to directly access someone's opinions without measuring them, we always need to choose a scale. And this is why we are focusing on how the choice of the scale may affect the model dynamics.

\subsection{Scales}

To avoid confusion in our upcoming analysis, it is necessary to explain what we mean by the word \textit{scale}. Researchers interested in the mathematical properties of scales, usually define them as a relationship between a construct and a subset of real numbers \citep{Eysenck}. Another (and perhaps more common) definition, relates to the process of building a scale. This definition states that a scale represents a set of items that can enable the quantification of a psychological construct. This last definition is the one that people have in mind when they refer to, for example, the Wechsler Adult Intelligence Scale (also known as WAIS). Besides the difference of focal point, the two definitions differ as the second may include finite precision and accuracy, while the first assumes perfect measurements.

An analysis of how measurement error impacts opinion dynamics models is beyond the scope of this article. Instead, we are interested in how choosing a relationship between the construct and numbers can strongly affect opinion dynamics models. Because of that, we will use the first definition, sometimes stressing that the analysis refers to the case of \textit{perfect accuracy and precision}.

\begin{figure}[!t]
\centering
\includegraphics[width=0.8\textwidth]{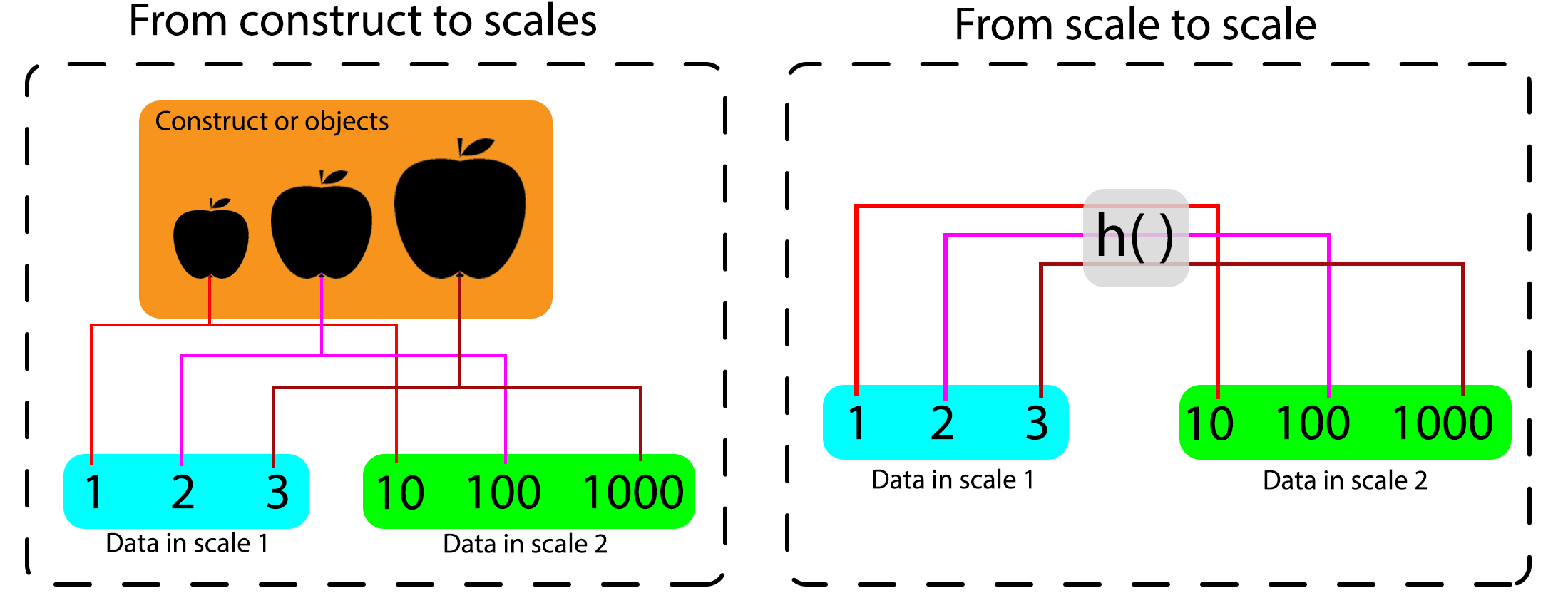}
\caption{(left) Schematic representation of the relationship between the levels of a construct (represented as apple size) and the numbers in two different scales. (right) Relationship between the two scales, represented as $h$}
\label{fig:X1}
\end{figure}

As we mentioned earlier, a scale with perfect precision maps each level of a construct to a single number. This means that level $l$ will be mapped to number $n_1$ in scale $S_1$ and to number $n_2$ in scale $S_2$. Consequently, a relationship exists between the two numbers, $n_1$ and $n_2$. In the following, we will call $h$ the relationship between the elements of two scales (see figure \ref{fig:X1} for a graphical representation) and we will write: $n_2 = h(n_1)$ (i.e. $h$ transforms $n_1$ into $n_2$). 

In statistical terms, this can be rephrased as the fact that Spearman’s rho correlation between the two scales is 1 (note that we are not using Pearson’s correlation as it would assume a linear relationship between scales). In terms of functions, we can say that a strictly monotonic function can be used to map one scale onto the other \citep{Tversky1}. This allows us to (1) discuss scales without referring them to a specific construct and (2) define a new scale $S_2$ just by using $h$, its relationship to  scale $S_1$ \citep{Tversky1, Tversky2, Tversky3}. 

Please, note also that a perfect scale should also have an infinite set of scores (or at least as many as the participants). This can be counter-intuitive for people involved in the practical use ordinal scales, as they usually have a really limited number of possible scores (e.g. a measurement scale comprising four items with five response options each only has 20 possible scale points). To understand why having a finite number of scores is not possible in perfect scales, just consider the case of rating 11 participants on a 10 point scale. This implies that at least 2 of them will be represented by the same score. If the strength of their opinion is not exactly the same, then the scoring will not be exact and the scale will not be perfect. This is why perfect scales should have an infinite number of possible scores within the defined extremes of the scale.

Sometimes in the text, we will also say that a model was \textit{seeded with a scale} or that \textit{we use a model with a scale} (e.g. "we use model $M_A$ with scale $S_1$"). This means that the construct was turned into numbers using the mentioned scale and that those numbers were used as input for the model.

\section{Theoretical background}
\subsection{Hypothesis on the opinion space}

In order to explore the concept of opinion space scientifically we need to introduce a testable hypothesis. As mentioned in the introduction, there have been no previous studies, to our knowledge, that analyze in depth the concept of opinion space, so we did not find any explicit claims about its properties. Because of that, here we propose two hypotheses which will help us eventually reveal hidden assumptions and, in general, better understand opinion dynamics models.

Hypothesis 1 (H1) is based on the idea that the choice of the scale should not affect the model dynamics. Formally, we write it as: \textit{All psychological scales measuring the same construct with infinite accuracy and precision produce equally valid results when seeded to the same opinion dynamics model.}

The practical meaning of H1 is that, whatever the scale, the model will produce the same result (we will properly define the concept of \textit{same result} in the next section). Or we can say that,  as long as the same construct is being (perfectly) measured from the same people, every scale can be used equivalently as model’s input. If H1 turns out to be false, it means that, even if the two scales are perfect, they will produce different results when used in the same model.

The second hypothesis is based on the idea that it should be possible to find a scale (such as interval scales) which should work with every model. This would need to be true in order to directly compare different models without discussing the properties of their opinion spaces, as is common practice in review  articles \citep{Castellano_2009, Flache}. Phrased differently, this hypothesis tests the idea that one opinion space can be used equivalently across multiple models. Thus, the first version of this hypothesis (H2.1) states that: \textit{“given two models $M_1$ and $M_2$, whose opinion spaces are defined on the same interval, and capable of producing results with the same precision, if scale $S$ works perfectly for the model $M_1$ then it also works perfectly for model $M_2$.”} 

We can also introduce a more general version of this hypothesis (H2.2) claiming that: \textit{“if scale $S$ works perfectly for model $M_1$ then it also works perfectly for model $M_2$, if the opinion spaces of the two models are defined on the same interval.”}

In the following, we will analyze both versions of H2, making clearer the difference between the two. For now, let us just notice that, if H2.1 and H2.2 turn out to be false, we could find a scale $S_1$ which is the best scale for model $M_1$, and a different scale $S_2$ which is the best scale for $M_2$ (even if they are perfectly measuring the same construct). In terms of opinion space, this would mean that the opinion spaces of the two models are different (meaning that they require different scales).


\subsection{Topological equivalence}

When discussing H1, we mentioned the idea of finding the same results while using two different scales on the same model but did not explain what we meant by \textit{the same result}. A simple (but wrong) way to express this concept, would be to expect the same numeric output in the two cases. To understand why this is a poor approach, just consider the case of a toy model which outputs the mean height of two objects. If we input $1 m$ and $2 m$ we obtain an average of $1.5 m$. While if we input the height of the same objects in feet, we obtain an average of $4.9 ft$. This last value, is \textit{equivalent} to the $1.5 m$ we just found but \textit{not numerically identical}.

What we have just shown is that, when changing scale, we do not expect to obtain the same numbers as output (indeed, $1.5 \neq 4.9$). However, we expect to have a way to convert both the input and the output from one scale to the other.

\begin{figure}[!t]
\centering
\includegraphics[width=0.5\textwidth]{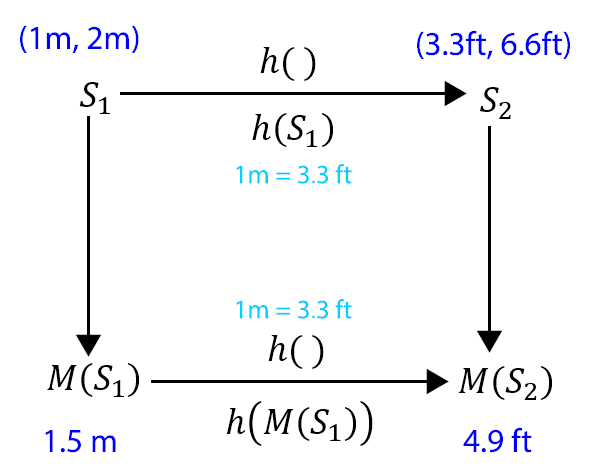}
\caption{A scheme of how $h$ converts $S_1$ to $S_2$ both at the input and at the output of the two model instances. The input in meters (top left) can be converted into feet (top right) using the conversion function (light blue). The model will output two different numbers depending on what we provided as input (either meters, left or feet, right). However, by using the same conversion function, we can transform one output into the other. Thus, it does not make any difference if we use the model with one scale or the other.}
\label{X3}
\end{figure}

In the same way, with opinion dynamics models we do not expect to obtain the same numeric output when working with two different scales. Instead, we expect to be able to convert the outputs from the model when using different scales (see figure \ref{X3}). Thus, making the choice of the scale independent from the meaning of the result. On a qualitative level, this can be seen as the fact that if the model predicts people converging to pro-Trump opinions when using scale $S_1$, we expect to make the same prediction with scale $S_2$ as well, even if this result is represented with different numbers.

To mathematically express this concept (and so make it scientifically testable) we use the concept of topological equivalence \citep{Alligood}. To avoid overloading this section, we suppose that the model is deterministic (thus without random variations). We also suppose that the model produces a single output, represented as $M(x)$, where $x$ is the input data. Since these requirements are extremely stringent, we will relax them in the next section, thus extending our analysis to all continuous opinion dynamics models.

To be precise, we should also distinguish between the concept of input data and the scale used to produce those data. For example, we could find that two people have an opinion respectively of $0.2$ and $0.6$ on a scale. In this case, the two numbers $0.2$ and $0.6$ would be the data, while the relationship between the construct and those numbers would be the scale. Since this paper assumes perfect measurement with no noise or measurement artifacts, this distinction will not improve clarity in the present analysis. Thus, for simplicity, we will use $S_i$ to mean both the scale and the data measured in scale $i$.

With these premises, we can write the topological equivalence between two models as the equivalence of the model’s output with the two scales as:

\begin{equation}
\label{e5}
M(\mathbf{S_2} )=h(M(\mathbf{S_1}))
\end{equation}

Where bold is used to represent vectors, $S_1$ is the data in scale $1$ and $S_2$ the same data measured in scale $2$, $M$ the function which produces the model’s output and $h$ is the (already mentioned) transformation from one scale to the other, such that:

\begin{equation}
\label{e6}
\mathbf{S_2}=h(\mathbf{S_1})
\end{equation}

Notice that the left term of \ref{e5} is the result of using the model with scale $2$ (in figure \ref{X3}, this would be moving from top to bottom on the right side). While the right term is obtained by using the model with scale $1$ and then converting the result using $h$ (thus moving from top to bottom on the left of figure \ref{X3} and then moving to the right). Finally, \ref{e6} represents how $S_2$ can be obtained from $S_1$ (from top left to top right in figure \ref{X3}).

Note also that we are using $h$ acting both on scalars and vectors. This is because for every $\mathbf{x}$  $\epsilon$  $R^N$, $h(x)$ is found as:

\begin{equation}
\label{e6.5}
h([x_1,x_2,…,x_N ])=[h(x_1 ),h(x_2 ),…,h(x_N )]  
\end{equation}

If equation \ref{e5} holds, we say that using scale $1$ is equivalent to using scale $2$ with model $M$. This means that we can convert both the model's input and output from one scale to the other, obtaining the same results (as we did with meters and feet in our toy model). 

In the next sections, we will compare also the case of running two different models, $M_1$ and $M_2$, on two different scales. In this case \ref{e5} can be generalized to:

\begin{equation}
\label{e7}
M_2(\mathbf{S_2} )=h(M_1(\mathbf{S_1}))
\end{equation}

If equation \ref{e7} holds, we say that using scale $1$ with model $M_1$ is equivalent to using scale $2$ with model $M_2$. 

To see how \ref{e5} works with a model, let us recall the example of the model of the mean height (represented in figure \ref{X3}). In this case we are using the same model with two different scales. Thus, we use equation \ref{e5} with $M(x)=mean(x)$. Furthermore, we have $h(x)= 3.3x$, which is the conversion between the two scales ($1 m = 3.3 ft$). If we substitute all of this into equation \ref{e5} we find the same values on the left and on the right hand side, meaning that using meters or feet is equivalent for this model. Also, this can be stated as the fact that, it does not matter if we make calculations in meters or feet, as we can always convert one into the other using $h$.

Coming back to the example of Trump-related opinions we suppose that the two opinions are $[0.4, 0.6]$ which produces an average of $0.5$. If we use $h(x)= 10x$, we will have opinions $[4,6]$, which produce an average of $5$. Since $h(0.5)= 5$, we see that we convert the two results into each other using $h$. Thus, we can use any of the two scales without affecting the result.

\section{Theoretical analysis}
Here we analyse under which cases equation \ref{e5} holds for opinion dynamics models. Thus, in which cases the scale transformation do not affect the model’s output. To make the analysis more sound we relax the hypothesis of the model being deterministic and producing a single number as output. This can be done by requiring $M$ to represent a single step of the model, instead of being the final output. This means that $M(x)$ is just the updated opinion value of one agent. For example, if at each step an agent updates her opinion with the average of all the other agents’ opinions, we will have $M(\textbf{x})=mean(\textbf{x})$, where $\textbf{x}$ denotes the vector containing the opinion of the agents selected for the interaction.

We can easily see that equation \ref{e5} does not hold for every possible function. We can verify this, for example, by choosing $M(x)=2x$ and $h(x)=x^2$. However, by assuming specific functions it is also possible to find conditions under which the equation holds. For example, by assuming both $h$ and $M$ as linear functions, the equation is satisfied. Another interesting case in which equation \ref{e5} is satisfied is when an agent copies the opinion of another agent (such as in the voter model or in the Axelrod model). In this case equation \ref{e5} is valid for both linear and nonlinear transformations.

In general, we could say that \textit{non-arithmetic models} (i.e. models which rely uniquely on copying) are not altered by scale transformations. Instead, \textit{arithmetic models} (i.e. models which are based on averaging or other arithmetic processes) may produce different results when their scale is transformed.

This means that if we want to know all the arithmetic models that satisfy equation \ref{e5} (and so are unaffected by scale transformations), we should analyze them individually. Since a review of every model is beyond the scope of this paper, here we focus on just three models, which have been presented by Hegselmann and Krause \citep{HK_avg}. This will be enough to test the hypotheses and, at the same time, provide some insight into the importance of scale selection in opinion dynamics modelling.

\subsection{Definition of the models}
In the models we are going to analyze every agent has an opinion $x$ defined by a scalar variable in the interval $[0, 1]$. Furthermore, the model is characterized by a confidence interval $\varepsilon$. At each step:

\begin{enumerate}
    \item An agent $i$ is selected
    \item From all her neighbors, only those within her confidence interval $\varepsilon$  are selected
    \item The agent updates her own opinion averaging the opinion of all her selected neighbors (including herself)
\end{enumerate}

In the original paper, these models have been studied on complete graphs in the case of synchronous update. In our case we keep the network topology, but we study the models in cases of both synchronous and asynchronous update. Indeed, synchronous modelling has been shown to produce artifacts in some models \citep{Galan}, and we want to make sure that our results are not due to artifacts.

The three models differ from each other by their averaging function, leading to the following three functions (one for each model):

\begin{equation}
\label{e8.1}
       M_A (x)=\frac{1}{n}\sum_i x_i 
\end{equation} 
\begin{equation}
\label{8.2}
    M_G (x)=\sqrt[n]{\prod x_i}
\end{equation} 
\begin{equation}
\label{8.3}
    M_H (x)=\frac{n}{\sum_i \frac{1}{x_i}}
\end{equation} 

As these functions are measurements of mean, their output number is always between the minimum and the maximum of the input data. 

To simplify the theoretical analysis, in this section we analyze the case in which the confidence interval is the maximum for all three models. Meaning that everyone can interact with everyone else. This simplification will be removed in section 5, where we will study these models via simulations using the complete model specified in the original paper.

As mentioned, the opinion space is defined as the interval between $0$ and $1$. However, below we will also study what happens when we expand this interval (e.g. between $1$ and $100$). This is another procedure that has been previously performed in literature \citep{Gargiulo} that is worth considering. Thus, here, we will use topological equivalence to explore why this procedure might produce different results.


\subsection{Analysis of H1}

H1 was based on the idea that using different scales on the same model will provide the same result. As previously discussed, when $M$ and $h$ are linear, this is true (equation \ref{e5} holds), while this may not be the case with nonlinear models. For example, let us define a scale $S_1$ defined over the interval $[0,1]$. Then, let us introduce $S_2$ defined by $h_{1\rightarrow 2} (x)=99x+1$, which maps the interval $[0,1]$ of $S_1$ to the interval $[1,100]$ of $S_2$. In this case equation \ref{e5} holds perfectly for $M_A$, while it does not for $M_G$. To quickly confirm this last point, just consider two test values on $S_1$: $0$ and $1$. By substituting them in equation \ref{e5}, for $M_A$ we obtain the same result on the left and on the right hand side. This is not true for $M_G$, where we find $1$ on the left hand side and $100$ on the right.

This means that, while data on linear models can be linearly rescaled (and so the interval can be linearly extended), this is not true for nonlinear models. Indeed, these models can produce different results when using linear transformations of the same scale as input. Just to stress this concept, imagine the case of some opinion data in a scale $1$ to $7$ and the problem of how to fit them into $M_G$. A possibility would be to linearly rescale the data between $0$ and $1$, to preserve the model's interval. Alternatively, we can preserve the data values and just run the model in the interval $1$ to $7$. While the two procedures would be equivalent on a linear model like $M_A$, they will produce different results on nonlinear models such as $M_G$.

However, it is important to notice that this does not mean that rescaling is problematic only for nonlinear models. Indeed, by using $S_3$, defined by the function $h_{1\rightarrow3} (x)=x^2$ even $M_A$ does not produce the same results. We can show this using two test numbers such as $1/4$ and $3/4$. This means that if we plan to use $M_A$ we can linearly rescale data, while any other transformation (such as compensation for skewness) will alter the model dynamics. 

Another important comment regards the choice of the scale. We know that ordinal scales can be thought as nonlinear transformation of each other (as shown in figure \ref{fig:X10}). Thus, using different scales will produce different results even on linear models. To go back to the example for Wada’s data, we could choose whether we want to represent them using the classical numbering of Likert scales (1 to 6) or the ratings of positivity from the YouGov study. As we will show in section 6, this choice would produce different results when we run the model and there is no easy way to say which choice is more appropriate. The only way would be to test them and compare the their predictions with the dynamics of the real world.


\subsection{Analysis of M2}

For the following analysis, we consider the ideal case in which we found a scale $S_1$ which works perfectly with $M_A$ (i.e. makes predictions which perfectly adhere to reality) and we wonder if the same scale can be used for all the other models. This would mean that the scale that works best for one model, also works best for all the other models. 

Just by knowing that $M_A (x)\neq M_G (x)$ we can already tell that the two models will produce different results when using the same scale, thus H2.2 is false. However, this may just mean that $M_A$ is correct, while $M_G$ makes poor predictions. To prove this false, in the next lines we will show that if $M_A$ works perfectly, then even $M_G$ works perfectly, but on a different scale.

Let us consider scale $S_4$ defined by the transformation $h_{1\rightarrow 4} (x)=e^x$. Since the exponential function transforms the sum into a product, it is possible to show that:

\begin{equation}
\label{e20}
M_G (h_{1\rightarrow 4} (S_1))= h_{1\rightarrow 4} (M_A (S_1))
\end{equation} 

Which can be rewritten as:

\begin{equation}
\label{eq21}
M_G (e^S )=e^{(M_A (S))}
\end{equation} 

This equivalence means that if we use $M_G$ with $S_4$ we will always obtain the same results as using $M_A$ with $S_1$. Furthermore, the same results can be obtained with $M_H$ using $h_{1 \rightarrow 5} (x)=1/x$ (i.e. using $M_H$ with $S_5$ is equivalent to using $M_A$ with $S_1$). 

This shows two things. The first is just that H2.1 is false: we cannot find a single scale which is the best for every model. The second important consequence of this results is that we can find three different models ($M_A$, $M_G$ and $M_H$) each one producing perfect results within its own scale ($S_1$, $S_4$ and $S_5$ respectively). This also means that $M_A$, $M_G$ and $M_H$ are the same model under a scale transformation.


\subsection{Model invariance}

As discussed before, linear models can be linearly rescaled, while nonlinear models, in general, cannot. We can see this applying a linear transformation on $M_A$, such as:

\begin{equation}
\label{eq22}
M_A (\alpha x+q)=\alpha M_A (x)+q
\end{equation} 

Which is just the application of equation \ref{e5} with $h(x) = \alpha x + q$ and allowed us to say that the extension of the opinion space in linear models (as done in \citep{Gargiulo}) is legitimate. While this linear rescaling does not work on nonlinear models, such as $M_G$, it does not mean that their interval cannot be extended or rescaled. Indeed, notice that

\begin{equation}
\label{eq23}
M_G (\beta x^n )=\beta [M_G (x)]^n
\end{equation} 

This means that $M_G$ is invariant to rescaling of the type $\beta x^n$. This appears more natural if we remember that $M_G$ is the exponential transformation of $M_A$. Thus, the linear rescaling of one model is transformed in the exponential rescaling of the other. This also solves the problem of how we can rescale the data for nonlinear models.

As we will discuss in the final section, this property may be useful not just to know how to rescale each model, but also for finding equivalence between different models.


\section{Analysis via agent-based simulations}
\subsection{Method}

In the previous sections we performed a theoretical analysis of the aforementioned Hegselmann Krause models. However, to simplify the analysis, we added some important constraints. The first one was performing the analysis using the maximum possible confidence interval. Even if this case is allowed by the original models, it is also a special case. Indeed, by allowing every agent to interact, we obtain a model with no real confidence interval.

The second major assumption stated that two models are considered equivalent if they produce the same result at every step. This could be considered a stringent requirement. Indeed, we may be interested just in the case where two models always produce the same output result, without any restriction on the dynamics.

Furthermore, in our previous analysis we often used just two numbers to test our hypotheses. However, in standard simulations we have (1) random fluctuations and (2) many more agents. So, it is possible that, when using a large number of agents, the effect of using different scales is completely hidden by the random fluctuations of the model itself. Another way to state this would be to notice that, even if two models are not mathematically identical, their difference during use may be so small as to be negligible.

To analyze this possibility and weaken our previous assumptions, we perform simulations on two of the previously mentioned models (specifically $M_A$ and $M_G$). To make sure that our results are not due to a small number of agents, we performed simulations with $N = 10,000$. This is way higher than most publications on agent-based models. Following the standard introduced by the authors of the model, we used a fully connected graph as topology and all agents were initialized with a random opinion from a uniform distribution. Hegselmann and Krause tested their models using synchronous updating, but, as this technique has been shown to produce artifacts in some models \citep{Galan}, we tested the case of both synchronous and asynchronous updating. The synchronous case was run with a maximum of $100$ iterations, each iteration updating all the $10,000$ agents. The asynchronous case was run $100*10,000$ iterations, each iteration updating only one random agent. As observed also by Hegselmann and Krause, this was more than enough for the model to converge. In order to collect enough random variations, we repeated each run $100$ times, each time collecting the average of the final distribution. As the average changed slightly each time (due to random variations), this resulted in a distribution of averages (see figure \ref{X5}).

For standard $M_A$ we used an opinion space between $0$ and $1$, and a symmetric confidence interval defined by $\varepsilon =0.2$. When we used scale transformation, we transformed both the space and the confidence interval, obtaining an asymmetric interval. This has been done as the confidence interval determines the agents that are selected for the interaction. By using this transformation, there is no difference in the selected agents when using each one of the scales.  Indeed, both the opinion space and the confidence interval should be measured with the same scale, thus it would not make sense to transform only one of them.

Agents’ opinions were initialized with random numbers from a uniform distribution. Scale transformations were applied after initialization.


\subsection{Results}

\begin{figure}[!t]
\centering
\includegraphics[width=0.6\textwidth]{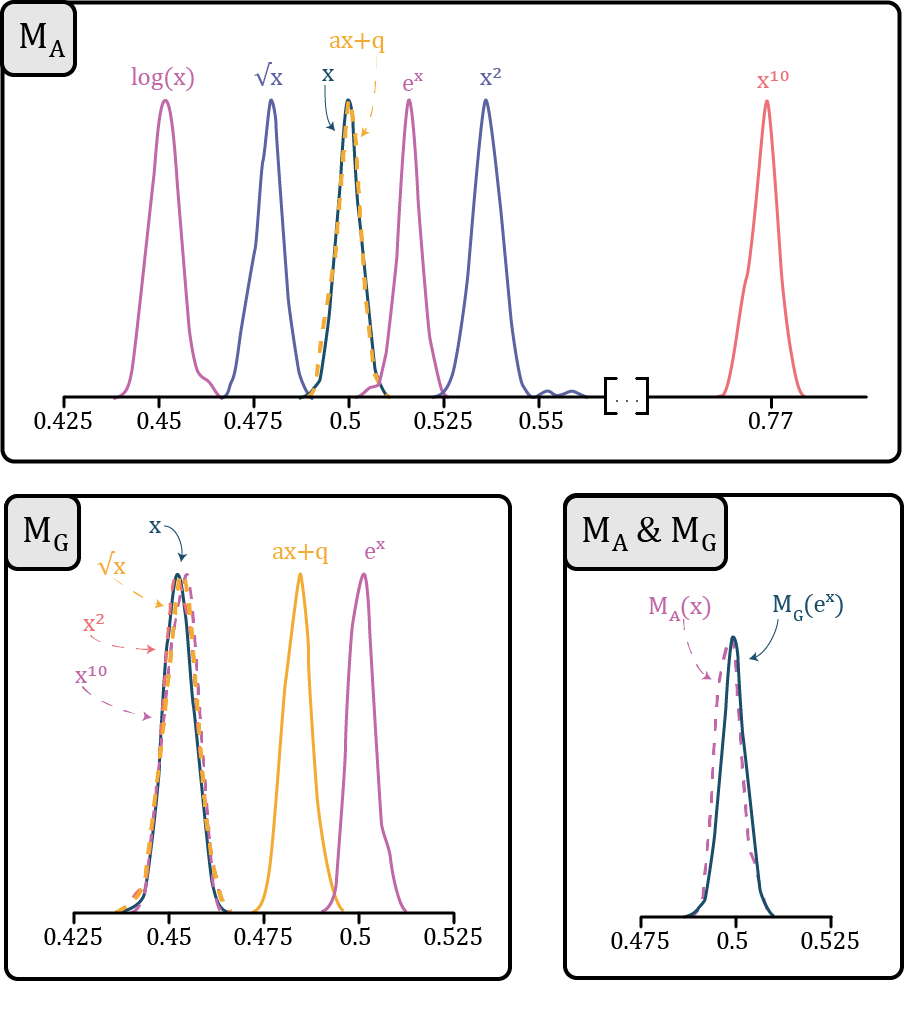}
\caption{Distributions of the average opinion. Each configuration (represented by a single curve) was repeated 100 times. Curves have been smoothed using a kernel density estimate. $M_A$ predictions are the same under linear rescaling, while they change for nonlinear transformations. $M_G$ instead changes when linearly rescaled, but is invariant to transformations of the type $x^n$. $M_A$ without scale transformation is equivalent to $M_G$ with the scale transformation $h=e^x$}
\label{X5}
\end{figure}

For each iteration, we collected the average of the final opinion distribution. In figure \ref{X5} we plotted the distribution of these averages when running the model $100$ times using the same parameters. Curves have been smoothed using the kernel density estimate provided by the Seaborn package for Python \citep{Seaborn}. Results were almost identical for the case of synchronous and asynchronous update, thus we decided to plot only the asynchronous case. To compare the results, we transformed each distribution using the inverse function (in agreement with the concept of topological equivalence, equation \ref{e5}). This allowed us to have all the final distributions in the $[0, 1]$ interval. Without this transformation each scale would be in a different interval (e.g. the logarithmic one would be in the interval $[1,e]$) making impossible to compare the final results. This process is further detailed in the next section.

To confirm the theoretical analysis and the idea of an asymmetric threshold, we run both $M_A$ with no scale transformation and $M_G$ with the exponential scale transformation. According to the theoretical analysis, these two models should be equivalent. As shown in figure \ref{X5}, the two curves overlap almost completely, in agreement with our expectations.

To observe the effects of different scales on $M_A$, we used five different transformations: linear, $x^2$, square root, exponential, and logarithmic. Curves show almost no overlap (thus producing different predictions) except in the case of the linear rescaling, which, in agreement with the theory, did not change the model’s output. We run a similar analysis also for $M_G$ showing that is indeed affected by linear rescaling but not from transformation of the kind $\alpha x^n$. Also this is in good agreement with our theory.

By looking at the graph of $M_A$, we could be tempted to state that the uncertainty on the scale leads to an error of roughly $\pm 5\%$. However, this result may change by using different transformations. For example, by looking at scale transformations of the type $x^n$ we can find the following limits for $M_A$:

\begin{equation}
\label{e10}
\lim_{n\rightarrow \infty} \left[ mean(\textbf{x}^n) \right]^{\frac{1}{n}} = \max{[\textbf{x}]}
\end{equation} 

\begin{equation}
\label{e11}
\lim_{n\rightarrow \infty} \left[ mean(\textbf{x}^{-M}) \right]^{\frac{-1}{n}} = \min{[\textbf{x}]}
\end{equation} 

The first equation means that as we use transformations with higher n (in the scale transformation $x^n$) the output of $M_A$ moves gradually from 0.5 to 1. We tested this assumption running $M_A$ with the transformation $x^{10}$ and obtaining an average of $0.76$ (thus an increase of roughly $50\%$ to the case of $M_A$ with no transformation). This is also shown if figure \ref{X5}. A similar result can be obtained using the transformation $x^{-n}$ which, according to equation \ref{e11}, will move the final average of the model from $0.5$ to $0$. This means that if we allow all possible scale transformations, our uncertainty would be of $\pm 100\%$ as the convergence number could be anywhere in the entire interval $[0, 1]$.

As we will outline better in the next section, it is true that we do not know which is the best scale for representing the data (in terms of opinion dynamics models) but it is also true that we will not expect to consider transformations such as $x^{100}$. Thus, we should be able to find some range of admissible transformations, which will also result in a limited range of uncertainty on our models. Identification of similar ranges is an entire new study and beyond the scope of this paper. Indeed, we expect it to depend on the model itself, as well as on the network topology which can amplify or reduce this uncertainty.

\subsection{Further clarification}

Some readers may think that the difference in the qualitative result is not surprising. Indeed, coming back to our Trump-related example, we may imagine to find an output average of $0.6$, thus predicting a pro-Trump opinion. Now, if we  consider the scale transformation $x^2$, we may think that the model should converge to $0.6^2=0.36$. As this is below $0.5$ this should predict an anti-Trump average opinion.

As mentioned in section 3, this line of thinking is wrong, or, at least, it is not what we studied in this paper. Indeed, when we transform from one scale to the other, we transform all the points (including the threshold separating pro-Trump from anti-Trump opinions). So, in this case, everything below $0.5^2 =0.25$ will predict a pro-Trump opinion.

To show this, let us imagine that the model ran with the second scale converged to 0.16. Since 0.16<0.25 we could already tell that here the model is converging to an anti-Trump opinion. However, to better compare this result with the first, we have to transform it back to the original scale. Since we squared the scale, to transform it back we have to use the inverse function, obtaining $\sqrt{0.16}=0.4$. Thus this (hypothetical) model will predict a result of 0.6 with the first scale and 0.4 with the second scale. 

\begin{figure}[!t]
\centering
\includegraphics[width=0.8\textwidth]{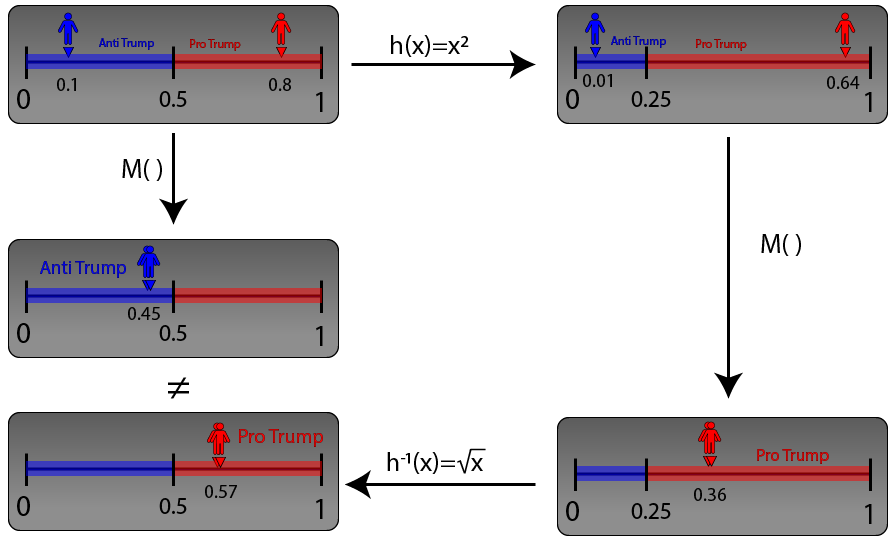}
\caption{Representation of how the the threshold is transformed together with the opinion transformations and how different simulations are compared.}
\label{X:trump_scheme}
\end{figure}


\section{Using real world scales}
So far we have analyzed the case of perfect scales. However, the choice of a scale affects not only the analysis of the model's properties, but also their applications to real data. To show this, here we analyze what happens when we run $M_A$ with two different scales which represent the same construct.

To start, imagine that we have collected some opinion data using the type of answers shown in Wada's study: [Very good, Good, Somewhat good, Somewhat Bad, Bad, Very bad]. Suppose also, that we found the following distribution [2\%, 21\%, 11\%, 29\%, 6\%, 31\%], which means that 2\% of people answered "Very good", 21\% answered "Good", etc. If we want to run an opinion dynamics model seeded with these data, we need to choose how to transform the answers to numbers (i.e. which scale to use). As noted before, we could either use standard Likert scale coding or an alternative one, such as the one obtained from the results of the YouGov study \citep{YouGov}.

If we suppose the maximal confidence interval ($\varepsilon = 1$) and we run $M_A$ on the Likert scale we obtain that everyone will converge to opinion: 4.09. Since this is not an interval scale, we cannot claim that our result is 0.09 above "Somewhat bad." Indeed, in ordinal scales we cannot say how much a value is bigger, but only if it bigger or smaller\footnote{Note that we could perform a similar operation in the previous sections (even if we were dealing with ordinal scales) as we were working with perfect scales and thus with infinite points. This means that every output number from the model was still one of the (infinite) numbers of the scale. This is not true with finite scales, where only a few numbers belong to the scale.}. However, we can still claim that this result lies between "Somewhat bad" (coded as 4) and "Bad" (coded as 5).

If now we run the same model using response option coding from the YouGov scale, we find that everyone ends up with an opinion of 5.48. In the YouGov scale, this value is between "Somewhat good" (3.16) and "Somewhat bad" (5.82), which is different from the previous interval.

This shows how the same construct from the same people may lead to different predictions when using two different scales (in this case, even using the same questionnaire). However, we still do not know how rare this situation is. In the next section, we show the probability of making different predictions while using these two different scales.

\subsection{The practical effect of real world scales}
In order to measure how rare is the case described in the previous section, we simulated different distributions. Each distribution was obtained using six random variables, each one representing the amount of people who selected one of the six answers (i.e. the first variable represented the fraction of people answering "Very good"). In this way we can simulate every possible distribution.

After, we initialized 1,000 agents accordingly to the previously obtained distribution. Two simulations were then run, one using conventional Likert scale and another using the scale from the YouGov study. Both scales were linearly renormalized between 0 and 1 before the simulation started. The two simulations used the same network (fully connected) and same confidence interval.

For each run we collected the interval of the final average (e.g. between "Good" and "Somewhat good") and the the number of clusters (e.g. 1 if everyone shared the same final opinion, 2 if two final opinions were obtained, etc). Finally, we compared how often using the two scales brought to the same results.

Table \ref{tab:1} shows that the two scales predict the same interval for the final average roughly 78\% of the times, almost independently from the confidence interval. The situation is more complex when we consider the number of clusters. Indeed, depending on $\varepsilon$ we could range from always obtaining the same result, to never. 

\begin{table}[h!]
\centering
\begin{tabular}{ | c | c | c | }
\hline
  & Same interval & Same num. of clusters \\ 
  \hline\hline
 $\varepsilon = 1 $ & 78\% & 100\% \\  \hline
 $\varepsilon = 0.4 $ & 77\% & 97\% \\  
 \hline
 $\varepsilon = 0.3 $ & 78\% & 46\% \\  
 \hline
 $\varepsilon = 0.2 $ & 79\% & 61\% \\  \hline
 
 $\varepsilon = 0.15 $ & 78\% & 0\% \\  
  \hline
\end{tabular}
  \caption{Percentage of times the model produced the same results with the two scales}
\label{tab:1}
\end{table}

This section shows practically that the use of different scales may have a serious impact not only on the properties of the model, but also on the meaning of its predictions. Notice also that here only two scales were studied, while almost infinite possibilities would be available. A full study of how different scales would affect $M_A$, or any other model, is out of the scope of this paper. Here we only wanted to show that this effect exists and that it is not negligible.

\section{Discussion and future studies}

Opinion dynamics models have the potential to be applied to everyday life, as they could allow to simulate the effect of different policies, like a digital twin of our society \citep{Saddik}. To reach such a result, better knowledge is required about how the models work and how they should be used with real world data. This paper addresses both of these two aspects.

Indeed, up to now modelers studied the effect of the \textit{network topology} and the \textit{interactions rules} to understand the model’s dynamics. Here we showed that a third degree of freedom is also present: \textit{the scale}. Here we showed how changing the scale we can (1) strongly change the model’s dynamics, up to completely changing the result (equivalent to an uncertainty of $\pm 100\%$), (2) turn one model into another (both of them in the standard literature) and (3) find how to rescale data for nonlinear models. This shines new light on the problems of (a) exploring models’ properties and (b) validating models against real data.

People interested in \textit{model validation} should notice that there is no reason to believe that the scales most commonly used in psychometrics will provide the best predictions for a specific model. Therefore, these scales, when used in opinion dynamics models, could be considered transformations of an unknown \textit{best scale} which should be found for each model. This problem could be solved in many ways. 

A first method would be to just fix a scale and test the models against it. This would change the problem of "finding the best scale given a model" to "finding the best model given a scale."  Another method would consist of looking for explicit or implicit model specifications. For example, the Deffuant model \citep{Deffuant2000} is such that people will always move to an average position. Meaning that if a person with opinion $0.5$ interacts with a person of opinion $1$, both of them will move the same amount (e.g. $0.1$) towards the average (i.e. $0.75$). Knowing this, it is possible (in theory) to design a Deffuant scale in which people, after interaction, move by the same amount. Furthermore, new models could be designed with stronger assumptions about the nature of the scale or the opinion distribution. This would help researchers to identify the right scale, or directly control this degree of freedom.

Alternatively, some researchers may prefer non-arithmetic models, (i.e. models which rely on copying other agent's opinion). Indeed, we showed how this rule satisfies equation \ref{e5} and thus is insensitive to scale transformation. One example is Axelrod’s model of cultural diffusion \citep{Axelrod}, which can also be extended to resemble bounded confidence models, using ordinal scales instead of nominal \citep{MacCarron}. 

People interested in the study of \textit{model properties} could characterize the models' invariances. Thus, which models can be linearly rescaled and which, instead, should use different rescaling, such as  exponential rescaling. Eventually, it may be possible to find models which cannot be rescaled, which will pose a new challenge on how to process data. Similarly, models like $M_G$, can be only exponentially rescaled, meaning that the entire set of data should be either positive or negative. This poses a problem regarding what to do with scales which originally include both positive and negative values (such as a range $[-10, 10]$).

Research should also focus also on how each model is affected differently by the scale uncertainty. As mentioned in section 5, allowing all possible scale transformations may produce an uncertainty of $\pm 100\%$. However, we do not expect to be so uncertain about the nature of the scale to consider transformations such as $x^{100}$, thus we should identify which scales are acceptable. A good starting point for understanding the range of uncertainty may be the data transformations which are routinely used with psychometrics data (e.g. linear, square root and logarithmic). Thus modelers may test these scale transformations and observe how much their model is sensitive to scale transformation. In the meanwhile, we should also identify the set of acceptable transformations, to better characterize this process.

Another possible focus would be on ordinal scales, such as Likert scales. These are the most common type of scales for measuring opinion and have been already been used with opinion dynamics models \citep{Duggins}. However, Likert scales are nonlinear representations of interval scales \citep{Tversky1}, thus it will be interesting to study the typical "amount of nonlinearity” and how this may impact the overall result (notice that for carrying out this analysis we should also define a measurement of nonlinearity). We already performed a quick analysis in section 6 for $M_A$, but different models and scales should also be tested. Especially, it would be interesting to know if it is better to use scales with a fewer or more response options in opinion dynamics models. A similar problem was already studied for psychometric purposes \citep{taherdoost}, but the best conditions for agent based models may be extremely different.

Another case of nonlinearity lies in the relationship between models. Indeed, we observed that $M_A$ and $M_G$ are the same under a scale transformation. This also opens the door for the study of models’ invariances. Indeed, while $M_A$ is invariant to linear rescaling, $M_G$ is invariant to exponential rescaling. These properties could be studied to better understand the model properties, equivalence to other models and the type of nonlinearity.

Another important aspect to study regards the finite precision and accuracy of the scales (e.g. noise). This problem has already been introduced by Edmonds \citep{Edmonds2004}, and further explored by Polhill \citep{Ghost}, who showed possible artifacts (such as the appearance of "ghost agents") when using floating point numbers. However, to now very little is known about the interaction between finite precision, network topology and update rule. Indeed, they could possibly interact with each other, amplifying or limiting the overall uncertainty.

Similarly to the effect of finite precision, data can also include biases which can affect the model. This has been observed in deep learning, where just the sampling of data is enough to introduce race-based or gender-based biases \citep{Kay_unequal}. Thus, when considering applications of opinion dynamics models to the real world, this possibility should also be considered.

Finally, data analysis may also have a key role  in the evolution of opinion dynamics models. Indeed, the field of psychometrics up to now has generally developed scales aimed at psychological research or clinical applications. The current paper shows a gap between psychometric best practice and agent based models, which eventually could be bridged by producing data which optimized for opinion dynamics models.

Here we showed how the choice of a scale for representing data can strongly affect the opinion dynamics. This effect is strong enough to completely change the model's output or transform one model into another. Thus, similarly to the update rule and the network topology, we think that the scale can be considered as a new degree of freedom for opinion dynamics models.

\section*{Acknowledgements}
We would like to thank James Gleeson, Padraig MacCarron, Paul J. Maher, 
Caoimhe O'Reilly, Kevin Burke and Susan Fennell from University of Limerick, Ireland for the fruitful discussions.

This preprint is part of a project which has received funding from the European Union’s Horizon 2020 research and innovation programme under the Marie Skłodowska-Curie grant agreement No 891347.




% ...





 
\bibliographystyle{jasss}
\bibliography{bibliography.bib} 


\end{document}